\documentclass[aps,pra,twocolumn,superscriptaddress,showpacs]{revtex4}
\usepackage{amsmath}
\usepackage{amssymb}
\usepackage{graphicx}
\usepackage{amsbsy}

\begin{document}

\title{Strongly localized moving discrete dissipative breather-solitons in Kerr
nonlinear media supported by intrinsic gain}

\author{Magnus Johansson}
\email{mjn@ifm.liu.se}
\homepage{https://people.ifm.liu.se/majoh}
\affiliation{Department of Physics, Chemistry and Biology (IFM), Link\"{o}ping
University, SE-581 83 Link\"{o}ping, Sweden}

\author{Jaroslaw E. Prilepsky}
\affiliation{Aston Institute of Photonic Technologies, Aston University,
 B4 7ET, Birmingham, UK}

\author{Stanislav A. Derevyanko}
\affiliation{Department of Physics of Complex Systems, Weizmann Institute of
Science, Rehovot 76100, Israel}

\begin{abstract}

We investigate the mobility of nonlinear localized modes in a generalized 
discrete Ginzburg-Landau type model, describing a one-dimensional
waveguide array in an active Kerr medium with intrinsic, saturable gain and
damping. It is shown that
exponentially localized, traveling discrete dissipative breather-solitons
may exist as stable attractors supported only by intrinsic properties of the
medium, i.e., in absence of any external field or symmetry-breaking
perturbations. Through an interplay by the gain and damping effects, the moving
soliton may overcome the Peierls-Nabarro barrier, present in the
corresponding conservative system, by self-induced time-periodic oscillations
of its power (norm) and energy (Hamiltonian), yielding exponential decays
to zero with different rates in the forward and backward directions. In
certain parameter windows, bistability appears between fast modes with
small oscillations, and slower, large-oscillation modes. The velocities and the
oscillation periods are typically related by lattice commensurability, and
exhibit period-doubling bifurcations to chaotically ``walking'' modes under
parameter variations. If the model is augmented by inter-site Kerr
nonlinearity,
thereby reducing the Peierls-Nabarro barrier of the conservative system, the
existence regime for moving solitons increases considerably, and a richer
scenario appears including Hopf-bifurcations to incommensurately moving
solutions and phase-locking intervals. Stable moving breathers also survive in
presence of weak disorder.

\end{abstract}


\maketitle

\section{Introduction}

The concept of dissipative solitons, being localized, dynamical objects with
non-trivial internal energy flows, which may exist in open, nonlinear systems
due to balance between gain, loss, dispersion/diffraction and nonlinearity,
is by now established in many branches of physics \cite{aa05,aa08},
and in particular in
optics where many applications have been devised \cite{afo09}. Although
originally conceived for spatially continuous systems, they have
lattice counterparts as discrete dissipative solitons/breathers, which also
have been discussed in a large number of physical contexts (see, e.g.,
the review \cite{fg08}).

In order to provide a feasible mechanism for transport of localized quantities
of energy, mobile dissipative solitons are highly important objects. While
their existence as exponentially localized modes for continuous systems is
well-known and not surprising (see, e.g., \cite{aa97}), the issue is more
delicate for lattices where, for the corresponding conservative systems,
the lattice can be viewed as inducing a periodic Peierls-Nabarro (PN)
potential,
which must be overcome during the motion. As a consequence, in a generic
conservative lattice system moving localized modes are not expected to
exist as exact solutions, since their motion in the PN potential causes
radiation to be emitted and thereby decay of a single moving soliton/breather
(see, e.g., \cite{reps} for discussion and further references, and
\cite{pel11} for a mathematical treatment). However,
in a dissipative environment with energy input, one might hope that a balance
(at least in average) between gain and loss could still be established to
sustain a moving discrete soliton indefinitely, and at the same time
damp out its radiation into a tail exponentially decaying to zero.

In fact, scenarios similar to that outlined above have already been observed
in several contexts, theoretically/numerically as well as experimentally.
Probably the first thorough numerical studies of mobile discrete breathers
in dissipative lattices were reported in a series of papers by the Zaragoza
group \cite{Floria1,Floria2,Floria3,Floria4}, for the damped and ac-driven
discrete sine-Gordon
(Frenkel-Kontorova) model. They were found to appear as dynamical attractors
for a rather large range of parameters, with phonon tails exponentially
decaying due to the damping, and asymmetric due to Doppler shifts of radiation
emitted in forward and backward directions. Two different types of moving
breathers, with a small regime of bistability, were discussed in
\cite{Floria1,Floria2,Floria3,Floria4}: ``induced fast'' breathers, generated
from superthreshold, symmetry-breaking dynamical perturbations of stable,
pinned breathers, and ``spontaneous slow'' breathers appearing from depinning
parametric instabilities of pinned, quasiperiodic breathers. An intermediate
regime with breathers moving in a seemingly random, diffusive way was also
identified. Other properties
of moving breathers described were the possibility for
formation of bound states, and regimes of mode locking of the breather
velocity at rational multiples of the driving frequency. However, one should
note that due to the spatially homogeneous driving force, the oscillation
amplitude of the breather tails does not decay to zero but to some constant
value, and thus for an infinite system all those excitations would have an
infinite energy.

In optics, discrete dissipative solitons are probably most
commonly discussed for arrays of coupled-waveguide resonators,
``discrete cavity solitons'' (DCSs) \cite{pel04}, for which the standard
coupled-mode
equations, in the case of Kerr nonlinearity, share many basic features with
the small-oscillation limit of the
damped-driven Frenkel-Kontorova model. In \cite{elk07} it was found that
resting DCSs could be made mobile by imposing a symmetry-breaking perturbation
on the system by inclining the holding beam, thereby introducing a
phase gradient for the effective driving force. Further numerical analysis
for a similar system with saturable nonlinearity \cite{yc11} also reached
the conclusion that, generally, a sufficiently large symmetry-breaking was
needed for a stable stationary soliton to move. This scenario should be
analogous to that of the ``induced fast'' breathers of
\cite{Floria1,Floria2,Floria3,Floria4}, except that the symmetry breaking
perturbation is applied to the physical system rather than to the initial
condition. A more recent work \cite{el13} also identifies ``spontaneously
walking'', as well as uniformly moving, DCSs in the absence of external
symmetry breaking, in regimes of Hopf instability of stationary DCSs. These
are thus analogous to the diffusively moving and ``spontaneous slow''
breathers, respectively, of \cite{Floria1,Floria2,Floria3,Floria4}.

Experimentally, moving discrete dissipative breathers have been observed in
damped-driven electrical lattices, one-dimensional (1D)
\cite{English08,English10}
as well as more recently two-dimensional (2D) \cite{English13}. In the 1D
lattice in \cite{English08}, traveling breathers locked to a uniform driver
were found as generic nonlinear excitations in certain parameter regimes, with
velocities precisely tunable by the driver amplitude and frequency. The
mechanism for mobility was explained in \cite{English10} in terms of a
self-induced dc-distortion ``propelling'' the breather. In 2D
\cite{English13}, an erratic motion was found, comparable to that
from \cite{Floria1,Floria2} discussed above.

However, in all above discussed cases, the moving discrete dissipative
solitons/breathers were supported by uniform external driving, leading
necessarily to modes with non-zero tails and infinite energy. By contrast,
in this work we address the question whether moving nonlinear
localized lattice modes may be
supported solely by an intrinsic gain mechanism, thereby allowing for
finite-energy solutions with tails decaying to zero. Although the question is
of generic nature, we will address it here in the context of a generalized
discrete
Ginzburg-Landau type model, introduced by Rozanov's group \cite{vrsfs08,kkr08}
to describe the
propagation of monochromatic radiation in a system of weakly coupled
single-mode active optical fibers, characterized by saturable amplification
and absorption \cite{rosanovbook} and a Kerr nonlinear refractive index.
The properties of stationary, strongly localized discrete solitons in this
system were studied
in \cite{kkr08}, generalizing earlier results obtained in a cubic-quintic
approximation \cite{el03,el05}. Indeed, also moving solitons were reported
for this system in \cite{vrsfs08}, but only for a special case with zero
Kerr nonlinearity, and with broad, continuum-like shapes. By contrast,
here we will mainly focus on showing that also strongly localized moving
solitons exist in presence of a standard, on-site Kerr nonlinearity, and
illustrate the mechanisms by which the interplay by gain and damping makes
it possible to overcome the Peierls-Nabarro barrier of the corresponding
conservative system, through self-induced time-periodic oscillations. It should
be noted that similar self-induced oscillations of a moving soliton were
recently also noticed \cite{kpd12}
in the transient dynamics for a continuous model with
linear gain and nonlinear loss and a spatially periodic modulation of the
linear as well as the Kerr nonlinear refractive index; however, due to the
linear gain the background of these solutions were unstable and therefore they
were not dynamical attractors.

In Sec.\ \ref{sec:model} we describe the generalized discrete Ginzburg-Landau
model of \cite{vrsfs08,kkr08}, which we also extend with possible inter-site
Kerr-nonlinearities and on-site disorder. We show and discuss our main
numerical results in Sec.\ \ref{sec:numerics}. Moving solitons for
the standard on-site Kerr nonlinearity are described in Sec.\ \ref{sec:Q0},
while additional features appearing in presence of inter-site nonlinearities
are discussed in   Sec.\ \ref{sec:Qne0}, and the survival of the moving
solitons also in presence of weak disorder is shown in
Sec.\ \ref{sec:disorder}.
Finally, some concluding remarks are made in Sec.\ \ref{sec:conclusions}.

\section{Model}
\label{sec:model}

The model to be used in this paper is a generalization of the model
from \cite{kkr08}, which we take in the following form:
\begin{eqnarray}
i \dot \psi_n + C(\psi_{n-1}+\psi_{n+1})
\nonumber \\
+\left( V_n + |\psi_n|^2  - i f_d(|\psi_n|^2) \right) \psi_n
+Q f_{is}(\{\psi_n\}) = 0 .
\label{master-eq}
\end{eqnarray}
Here, the dot will be referred to as a time-derivative, although evidently
it should be interpreted as a derivative with respect to a longitudinal
coordinate $z$ for the physical system of coupled identical
fibers in \cite{kkr08},
describing discrete spatial optical solitons. In this model, $\psi_n$ is the
mode amplitude in the $n$th fiber, and $C$ is the coupling constant between
neighboring fibers, which we here take to be real and thus neglecting losses
in the medium between the fibers (in \cite{vrsfs08,kkr08}, possible absorption 
or gain in this medium was also considered by adding a small imaginary part 
to $C$). We have also included a linear, real
on-site potential $V_n$ to allow for possible disorder in the fiber properties
(although we will take  $V_n \equiv 0$ for most of the paper).
The real function $f_d(x)$ describes the amplification and
absorption characteristics of each fiber, and is taken as in
\cite{vrsfs08,kkr08,rosanovbook}
to have the form
\begin{equation}
f_d(x) = - \delta + \frac{g}{1+x}  - \frac{a}{1+bx} ,\quad \delta, g, a, b >0 .
\label{fd}
\end{equation}
Here, $\delta$ describes the linear, non-resonant losses, $g$ and $a$ the
strength of the saturable gain and absorption, respectively, and $b$ the
ratio between the gain and absorption saturation intensities. 
We have chosen to normalize the field according to the gain saturation intensity and without loss of generality assumed that the Kerr coefficient is equal to unity.
A related model with purely
linear losses ($a=0$) was also studied in \cite{pyjd12}
in the context of DCSs (i.e., with
an additional constant term in the right-hand side describing the amplitude
of a holding beam), where also real-world parameter estimates for coupled GaAs
semiconductor waveguide resonators were given.

In addition, we have augmented Eq.\ (\ref{master-eq}) with a term describing
a possible inter-site Kerr nonlinear refractive index with a strength $Q$,
of the form
\begin{eqnarray}
f_{is}(\{\psi_n\}) = 2 \psi_n \left(|\psi_{n+1}|^2 + |\psi_{n-1}|^2 \right)
\nonumber \\
+ \psi_n^* \left(\psi_{n+1}^2 + \psi_{n-1}^2 \right)
\nonumber \\
+ 2|\psi_{n}|^2 \left(\psi_{n+1} + \psi_{n-1} \right)
+ \psi_{n}^2 \left(\psi_{n+1}^* + \psi_{n-1}^* \right)
\nonumber\\
+ |\psi_{n+1}|^2\psi_{n+1}  + |\psi_{n-1}|^2\psi_{n-1} .
\label{is}
\end{eqnarray}
The form of this term follows from coupled-mode analysis of a
waveguide array embedded in a nonlinear Kerr medium \cite{oje03}; for
simplicity we consider here a special case of the two-parameter form derived
in \cite{oje03} by equating the parameters to the single parameter $Q$.
As shown in \cite{MJ06}, the conservative version of Eq.\ (\ref{master-eq})
($f_d \equiv 0$)
also appears as a rotating-wave type approximation of a chain of anharmonic
oscillators coupled with anharmonic inter-site potentials,
with $Q=0$ corresponding to pure on-site (Klein-Gordon chain) and $Q=1/2$
to pure inter-site (Fermi-Pasta-Ulam chain) anharmonicity. The reason
for incorporating $f_{is}$ into our model is, that this type of inter-site
nonlinearity may drastically reduce the Peierls-Nabarro potential for the
conservative model (in the two-parameter model it may even be strictly zero
at special parameter values \cite{oje03}), and therefore highly facilitate
the movement of even strongly localized solutions \cite{oje03,MJ06}. As we will
see later, the term $f_{is}$ is not necessary for finding strongly localized
moving solutions as dynamical attractors to  Eq.\ (\ref{master-eq}) (and
indeed we will put $Q=0$ for the main part of this paper); however a nonzero
$Q$ increases considerably their existence region and also allows for a
richer dynamical scenario.

In the conservative case $f_d \equiv 0$, the two conserved quantities of
Eq.\ (\ref{master-eq}) are the total power (norm),
\begin{equation}
P=\sum_{n} |\psi_{n}|^2 ,
\label{norm}
\end{equation}
and the total energy (Hamiltonian) \cite{oje03},
which for the general case with
nonzero $V_n$ and $Q$ can be expressed most conveniently on the
form \cite{MJ06}
\begin{eqnarray}
H =  \sum_n \left\{2C \sqrt{A_n A_{n+1}} \cos(\phi_n-\phi_{n+1})
+ \frac{A_n^2}{2}
\right. 
\nonumber\\
\left.
 + Q \sqrt{A_n A_{n+1}} 
\left[\sqrt{A_n A_{n+1}}
\left( 2 + \cos 2(\phi_n-\phi_{n+1})\right)
\right.\right.
\nonumber\\
\left.
\left.
+ 2 (A_n + A_{n+1}) \cos(\phi_n-\phi_{n+1})
\right]  + V_n A_n
\right\} ,
\label{ham}
\end{eqnarray}
where we have expressed the complex mode amplitudes in terms of real
action-angle coordinates, $\psi_n = \sqrt{A_n} e^{i \phi_n} $. In the
``standard'' definition of a Peierls-Nabarro (PN) barrier
\cite{Eilbeck86,kc93}, one imagines a
localized mode sliding with small velocity across the lattice, and the PN
barrier is then defined as the difference in $H$, at fixed $P$, between
the two stationary modes centered at, respectively in-between, lattice sites.
In typical situations, one of these modes is stable and the other unstable,
and therefore they usually correspond to min and max, respectively, of an
imagined PN potential. However, in some models (and notably the model
(\ref{master-eq}) with nonzero $Q$ \cite{oje03}),
these two modes exchange stability
through symmetry-breaking bifurcations under parameter variation, and there
exists a small parameter regime of existence of additional, symmetry-broken,
stationary solutions \cite{aub06,oje03}
which also must be taken into account when determining
the ``true'' PN-barrier.

As explained in \cite{kkr08}, a necessary (but by no means sufficient)
condition to have nontrivial localized solutions for the fully dissipative
model (\ref{master-eq}) can be obtained as a restriction of the gain
parameter to an interval,
\begin{equation}
\delta + \frac{a - \delta + 2 \sqrt {a \delta (b-1)}}{b}\leq g \leq \delta +a .
\label{ginterval}
\end{equation}
The upper limit follows from the condition of existence of a stable
zero-amplitude tail, and the lower limit from the condition of existence
of a solution with non-zero amplitude for which the dissipative function
$f_d(|\psi_n|^2)$ in (\ref{fd}) is not always negative. Notice also
that the condition to have a nonvanishing interval in (\ref{ginterval}) imposes
an additional restriction on the damping parameters $\delta, a, b$, which
can be expressed as
\begin{equation}
\delta \leq a (b-1) .
\label{dampcond}
\end{equation}
Let us finally in this section also remark, that expanding the dissipative
function in (\ref{fd}) for small $x$ yields (in a cubic-quintic approximation
for $x=|\psi_n|^2$ such as in \cite{el03,el05})
\begin{equation}
f_d(x) \simeq g - \delta - a + (g - ab) x + (g - ab^2) x^2 ...
\label{3-5}
\end{equation}
However, as we will see, the relevant solutions describing moving localized
modes are typically found with peak values of the order of unity or larger,
and thus result essentially from the strong saturabilities of the
gain and damping parts of the dissipative function on different intensity
scales, and cannot be described by the cubic-quintic approximation (\ref{3-5}).

\section{Numerical results}
\label{sec:numerics}

In order to systemize and get out the most essential features of the
moving localized solutions found in certain regimes of parameter space,
we will from now on fix the damping parameters in (\ref{fd}) 
to have the same values as those
used for the study of stationary localized modes in \cite{kkr08}
(see also \cite{vrsfs08,rosanov05} and references therein),
namely
\begin{equation}
\delta=1, a=2, b=10 .
\label{parameters}
\end{equation}
Numerically, the largest possible existence regime (\ref{ginterval}) then
becomes $1.94853 < g < 3$. Notably, the large value of $b$ implies that the
absorption saturates at a much
lower intensity than the gain. We have checked that these parameter values
are not exceptional, e.g., by varying $\delta$ up to 1.5, $a$ down to 0.5, and
$b$ up to 20. Although evidently quantitative features change, the qualitative
picture remains similar as long as the possible existence regime determined
by (\ref{ginterval})-(\ref{dampcond}) remains non-negligible. Unless
otherwise noted we will assume $V_n \equiv 0$. We first (Sec.\ \ref{sec:Q0}) describe
the scenario observed for pure on-site nonlinearity ($Q=0$) when varying
the gain amplitude $g$ and coupling constant $C$, and then
(Sec.\ \ref{sec:Qne0}) discuss additional features appearing in the
presence of inter-site nonlinearities   ($Q\neq 0$).

In order to 'seed' a moving localized solution, an initial Gaussian pulse
with an imposed phase-gradient was used as a trial solution. By properly
adjusting parameters, such pulses were found by direct time-integration
of the equations of motion to belong to the basin of
attraction of exact moving localized solutions with specific velocities
$v$. Once found, these solutions were used as new initial conditions in
continuation versus the parameters $g, C, Q$. (In practice, since the
existence regime of moving solutions is larger in presence of inter-site
nonlinearities, it turned out
to be easier to first find moving solutions for some nonzero $Q$, which then
could be continued back to the, possibly more physically interesting, case
$Q=0$.)

The velocities were determined directly as
$v=\Delta n_{max}/\Delta t$, with $\Delta n_{max}$ being the translation
of the soliton peak during a (large) time $\Delta t$,
disregarding the initial transient before
reaching the attracting state. As we will see below, a moving soliton
generally exhibits an internal oscillation frequency of its intensity
$A_n \equiv  |\psi_n|^2 $, yielding oscillations also in the quantities
$P$ and $H$, thereby allowing the PN-barrier of the corresponding conservative
system to be overcome. Assuming a single internal frequency $\omega_b$, a
moving soliton can be assumed to have a form analogous to a moving
discrete breather as defined in \cite{fk99} (see also \cite{pdcret}),
\begin{equation}
A_n(t) = A(\omega_b t, n - vt),
\label{mdb}
\end{equation}
with $A(x,y)$ being a function which is $2\pi$-periodic with respect to its
first argument and localized with respect to the second. If, for some
integers $p,q$, the internal frequency and the velocity are related by
a commensurability relation,
\begin{equation}
p \cdot 2\pi v = q \cdot \omega_b ,
\label{pq}
\end{equation}
 the solution will return to its initial shape after a time
$T= p \cdot 2 \pi/\omega_b$, translated $q$ sites. As we discuss
below, in the case $Q=0$ with pure on-site nonlinearity, apparently the
pinning potential is strong enough to always force such a commensurability
(in most cases with $p=q=1$). On the other hand, for nonzero $Q$ the
weakening of the PN potential also allows for incommensurate regimes
($p,q\rightarrow \infty$),  where the solution, although moving with a
perfectly well-defined velocity $v$,
 never exactly returns to
its initial shape (see Sec.\ \ref{sec:Qne0}).

\subsection{The case of on-site nonlinearity, $Q=0$}
\label{sec:Q0}
\begin{figure}[t]
\centering
\includegraphics[height=0.5\textwidth,angle=270]{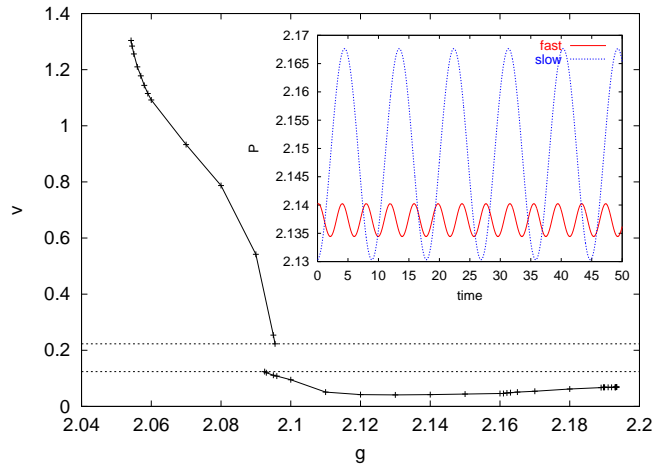}
\caption{(Color online) Main figure: velocity $v$ versus gain parameter $g$ 
for moving localized solutions
of Eq.\ (\ref{master-eq}) with $C=1$. Other parameters are
$V_n = Q =0$ and $f_d$ given by (\ref{fd}) with (\ref{parameters}). Horizontal 
dashed lines indicate the velocity gap. Inset:
Power (\ref{norm}) versus time for the bistable fast (solid red, small
oscillations) and
slow (dashed blue, large oscillations) solutions having the same $g=2.095$.}
\label{fig:Q0vel}
\end{figure}

Typical results for the variation of the soliton velocity with the
gain parameter $g$ in this case are shown in Fig.\ \ref{fig:Q0vel}.
For this particular value of the coupling constant, $C=1$, moving solutions
with well-defined velocities $v$ are found for
$2.0542 \lesssim g \lesssim 2.1935$. Notice that this interval is only about
13\% of the interval determined by the necessary condition (\ref{ginterval}).
The solutions are divided into two branches: rapidly moving solutions
for weaker gain, and slower solutions for larger gain (although the velocity
is not a monotonic function of $g$ for the latter branch). There is
a small regime of bistability, $2.0925 \lesssim g \lesssim 2.0955$, where
both solutions exist with different velocities. Notice also the ``forbidden
gap'' of velocities in the interval  $0.124 \lesssim v \lesssim 0.223$.
In this aspect, the scenario is thus similar to that for the
Frenkel-Kontorova model with ``induced fast'' and ``spontaneous slow'' moving
breathers \cite{Floria1,Floria2}.

Generally, as mentioned above, the moving solutions are associated
with time-periodic oscillations in $P$ and $H$, and in particular the slowly
moving solutions are associated with considerably larger oscillations as
well as larger average values of the power $P$ (see the inset in
Fig.\ \ref{fig:Q0vel} for comparison of two solutions for the same $g$
in the bistability regime).

\begin{figure}[t]
\centering
\includegraphics[height=0.5\textwidth,angle=270]{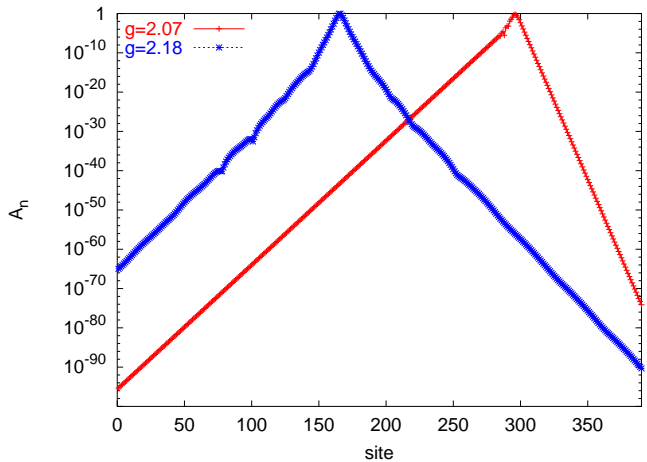}
\caption{(Color online) Snap shots of intensity $A_n$ in logarithmic scale for 
two right-moving solutions
with  $g=2.07$ (red, right peak, fast solution) and $g=2.18$
(blue, left peak, slow solution), respectively.
Other parameters are the same as in Fig.\ \ref{fig:Q0vel}.}
\label{fig:tails}
\end{figure}

In Fig.\ \ref{fig:tails} we compare two examples of snapshots of slowly and
rapidly moving solutions. Note that while both solutions are evidently
exponentially localized, the tails of the rapidly moving solution show
a much stronger asymmetry than those of the slow mode, with a much stronger
decay in the forward than in the backward direction. This can be
related to the Doppler shifts of emitted radiation as discussed in
\cite{Floria1}. For the slower solution, the velocity is $v\approx 0.06$ so the
effect is hardly noticeable, while the rapid solution moves about 15 times
faster, $v\approx 0.9$. Note also the crossover from one exponential decay
around the soliton center, to another, generally weaker, decay in the tails.

As seen in Fig.\ \ref{fig:Q0vel}, decreasing the gain on the
branch of rapid solutions leads to larger velocities, until a maximum velocity
$v \approx 1.304$ is reached for $g \approx 2.0542$. With a further decrease
of $g$, the solution will decay to zero after some time.

%
\begin{figure}[t]
\centering
\includegraphics[height=0.5\textwidth,angle=270]{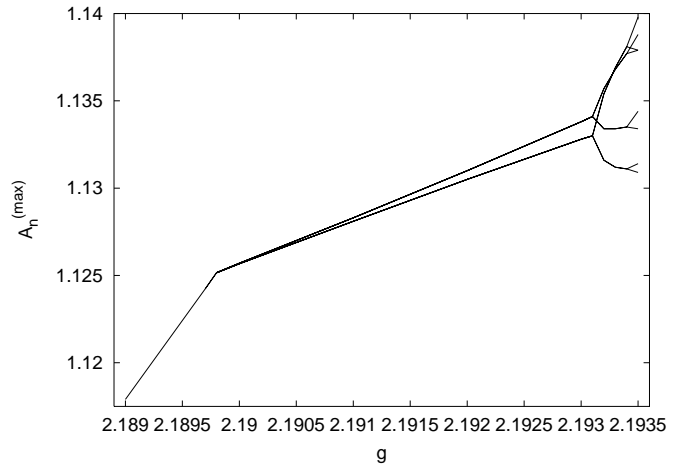}
\caption{Maximum intensity $A_n^{(max)}$ versus gain parameter $g$ for
different lattice sites, illustrating the locking between internal oscillations
and lattice translations for steadily moving solutions,
and the period-doubling scenario
in the upper end of the existence regime.
Parameters values are the same as in Fig.\ \ref{fig:Q0vel}.}
\label{fig:Amax}
\end{figure}

On the other hand,
increasing the gain on the branch of slow solutions leads to a more complicated
scenario. The power oscillation amplitude will increase, as well as its
average value, and develop strongly anharmonic features in contrast to the
nearly harmonic oscillations shown for smaller gain in Fig.\ \ref{fig:Q0vel}.
For $g\gtrsim 2.13$ the oscillations become double-peaked. While, for the
case $C=1$ shown in  Fig.\ \ref{fig:Q0vel} the oscillations are 1:1-locked with
the lattice translations [$(p,q)=(1,1)$ in Eq.\ (\ref{pq})]
in most of the existence
regime for the moving solutions, a period-doubling scenario is observed
when increasing $g$ even further. Here, we found a solution with period-2
[$(p,q)=(1,2)$] in an interval $2.1898 \lesssim g \lesssim 2.1931$,
period-4 [$(p,q)=(1,4)$] when $2.1932 \lesssim g \lesssim 2.1934$,
period-8 [$(p,q)=(1,8)$] in a small interval around $g\approx 2.1935$, etc.
The period-doubling scenario is illustrated in Fig.\ \ref{fig:Amax}.
For a further increase of $g$ the power oscillations become irregular,
resulting in an apparently chaotically moving solution which, similar to
the examples shown in \cite{Floria1,el13}, may move with almost constant
velocities for some time, then get trapped for some brief interval, after
which it restarts its motion in an apparently random direction.
For these parameter values, the randomly walking solution appears only in
a small interval of $g$, and a further increase ($g \gtrsim 2.205$) yields a
splitting of the soliton into two oppositely moving fronts
(``switching waves'' \cite{rosanovbook,kkr08})
so that all lattice sites will end
up at a constant, nonzero, intensity.

%
\begin{figure}[t]
\centering
\includegraphics[height=0.5\textwidth,angle=270]{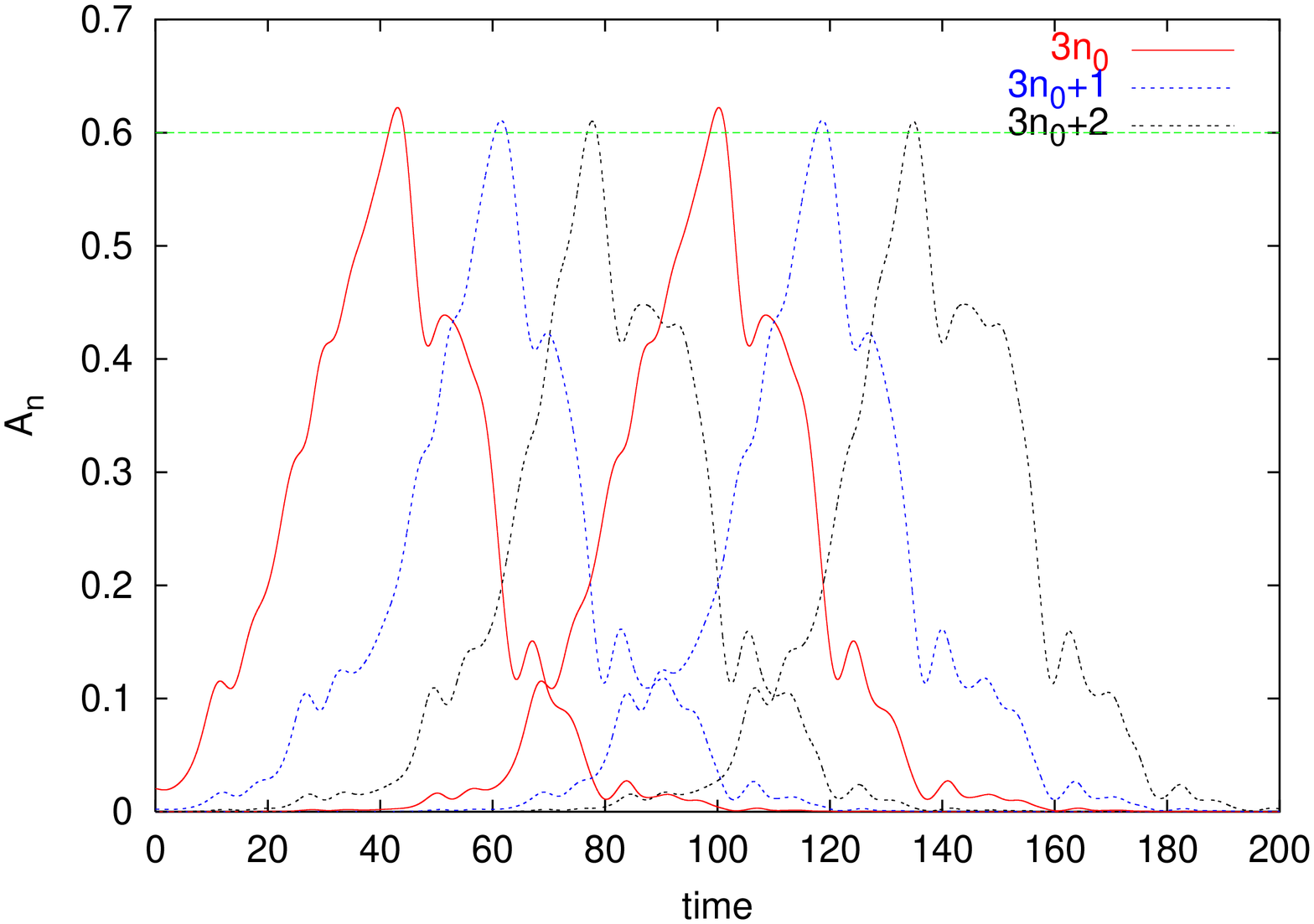}
\includegraphics[height=0.5\textwidth,angle=270]{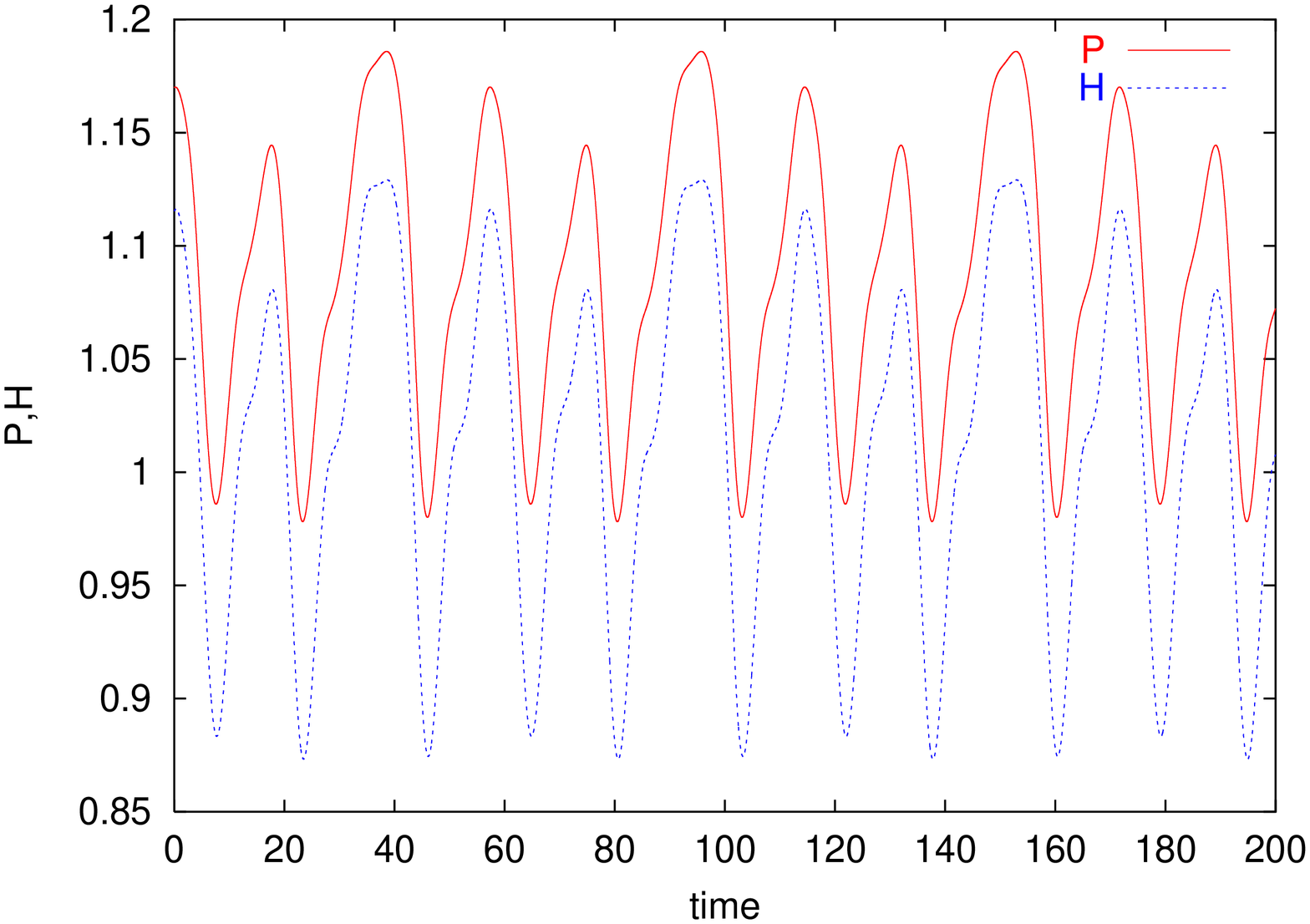}
\caption{(Color online) 
Dynamics of a moving soliton with velocity $v\approx 0.052$
in the period-3 window [$(p,q)=(1,3)$]
for $C=0.46$ and $g=2.0574$
(other parameters same as in Fig.\ \ref{fig:Q0vel}).
Upper figure: intensity  $A_n$ versus time
for six consecutive sites (larger $n$ corresponds to later peak time).
Note the identical curve shapes for every
third site (the horizontal line at 0.6 is a guide to the eye).
Lower figure: the corresponding oscillations of the quantities
$P$ (\ref{norm}) (solid red line) and $H$ (\ref{ham}) (dashed blue line). }
\label{fig:period-3}
\end{figure}

With the decrease of the coupling constant $C$, the range of $g$ where steadily
moving localized solitons exist narrows, and for $C \lesssim 0.45$ we were
not able to find any moving solitons at all. Moreover, all solitons generally
get slower when $C$ decreases, which is consistent with the interpretation
of a larger PN-barrier for strongly discrete lattices. In particular, for
$C=0.46$ (close to the lower existence boundary) we observe moving solitons
for $2.0528 \lesssim g \lesssim 2.0602$ with velocities
$0.045 \lesssim v \lesssim 0.064$ (monotonously decreasing with $g$). Also
here there are two main branches of solution as for $C=1$, however without
multistability. Instead, the ``fast'' branch now also exhibits period-doubling
bifurcations into a regime of chaotically moving solutions starting at
$g \approx 2.0564$, while the ``slow'' branch first appears at
$g\approx 2.0600$. Interestingly, in addition to the period-doublings
there are also windows of higher-periodic regularly moving solutions
inside the regime of chaotic solutions, analogously
to the standard Feigenbaum scenario. For example, we found a period-3 window
[$(p,q)=(1,3)$] for $2.0569 \lesssim g \lesssim 2.0574$, bifurcating
to period-6 [$(p,q)=(1,6)$] around $g\approx 2.0575$, etc.
The dynamics of a stable moving period-3 solution is illustrated in
Fig. \ref{fig:period-3}. Note that even though three different types
of peaks are clearly visible in the (strongly anharmonic)
oscillations of power and Hamiltonian,
the resulting differences in local intensity peak heights are hardly
distinguished on the scale of Fig.\ \ref{fig:period-3}.

\subsection{The case of intersite nonlinearity, $Q\neq 0$}
\label{sec:Qne0}

As mentioned above, including a nonzero $Q$ with an inter-site
nonlinearity (\ref{is}) for the conservative system (\ref{master-eq})
($V_n \equiv 0, f_d \equiv 0$) effectively reduces the PN-barrier and enhances
mobility of localized modes. The PN-barrier is expected to be smallest in
the regime where the site-centered and bond-centered solutions exchange
stability; the exact location of this exchange depends slightly on the ratio
$P/C$ so that the upper limit for stability of the site-centered solution
is $Q_c \approx 0.25$ for $P/C \lesssim 1$, and
$Q_c \approx 0.26$ for $P/C \gtrsim 10$ (cf.\ \cite{oje03,jj13}).

\begin{figure}[htbp]
\centering
\includegraphics[height=0.5\textwidth,angle=270]{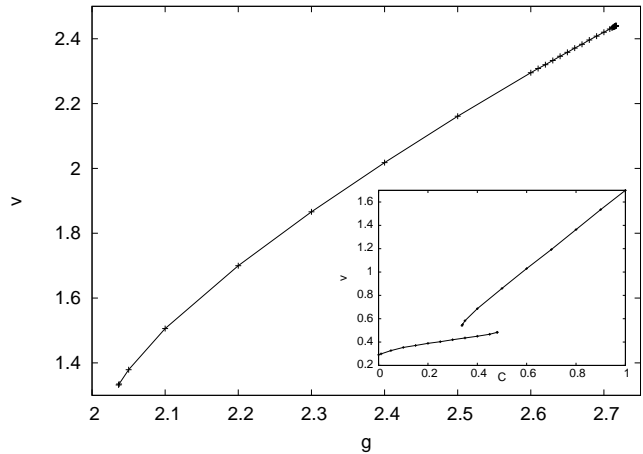}
\caption{Main figure: velocity $v$ versus gain parameter $g$ for moving
localized solutions
of Eq.\ (\ref{master-eq}) with $C=1$, $Q=0.4$, and other parameters
same as in Fig.\ \ref{fig:Q0vel}. Inset: velocity versus coupling constant
$C$ for fixed gain parameter $g=2.2$ and other parameters same as in main
figure.}
\label{fig:Q0.4vel}
\end{figure}

For the dissipative system with parameter values as in (\ref{parameters}),
we correspondingly observe a considerable enhancement in the existence regimes
for moving solitons when $Q \gtrsim 0.2$. So, for example, taking $Q=0.4$
and $C=1$
we find a continuous branch of rapidly moving solutions with small-amplitude
oscillations in the full interval
$2.0363 \lesssim g \lesssim 2.7166$, with velocities monotonously increasing
with $g$ from $v\approx 1.332$ to   $v\approx 2.440$
(Fig.\ \ref{fig:Q0.4vel}). Thus, compared to the
case $Q=0$ of Fig.\ \ref{fig:Q0vel}, the existence interval in $g$ is not
only 5 times larger, but also the dependence $v(g)$ is the opposite
(w.r.t. the fast branch).

Also the lower existence boundary in $C$ for moving solitons becomes
considerably smaller when $Q \ne 0$. Taking again, as an example,
$Q=0.4$ and fixing $g=2.2$, we find (see inset in Fig.\ \ref{fig:Q0.4vel})
that although the branch of stable rapidly
moving solutions continued from $C=1$ terminates at $C\approx 0.338$, there
is another branch of slow, large-amplitude oscillation solutions appearing
at $C\approx 0.481$ (i.e., again with an interval of bistability), which
can be continued all the way down to $C=0$. In this limit, the solution
moves with nonzero velocity ($v\approx 0.291$) solely due to the influence
of the nonlinear inter-site couplings.

\begin{figure}[htbp]
\centering
\includegraphics[height=0.5\textwidth,angle=270]{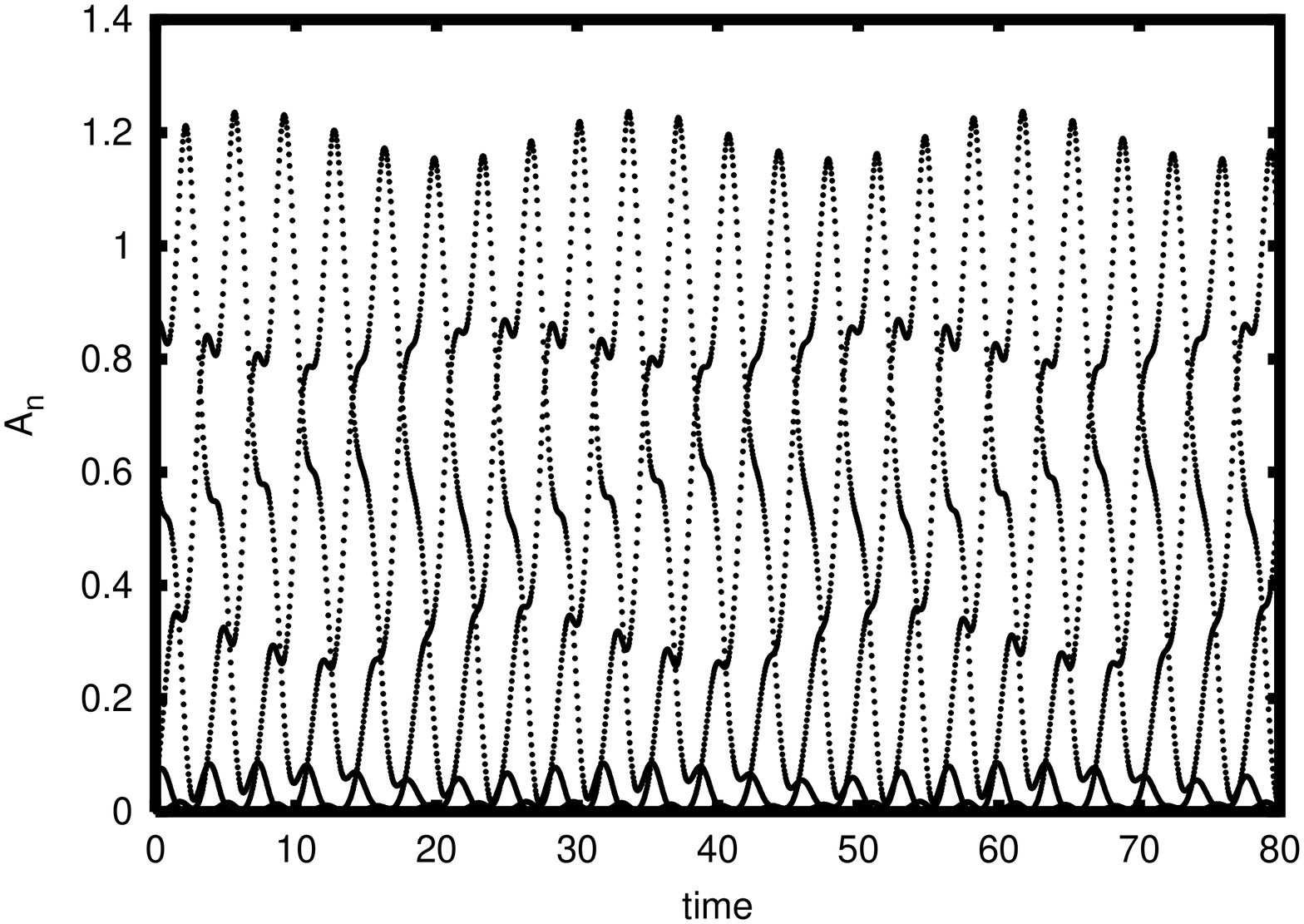}
\includegraphics[height=0.5\textwidth,angle=270]{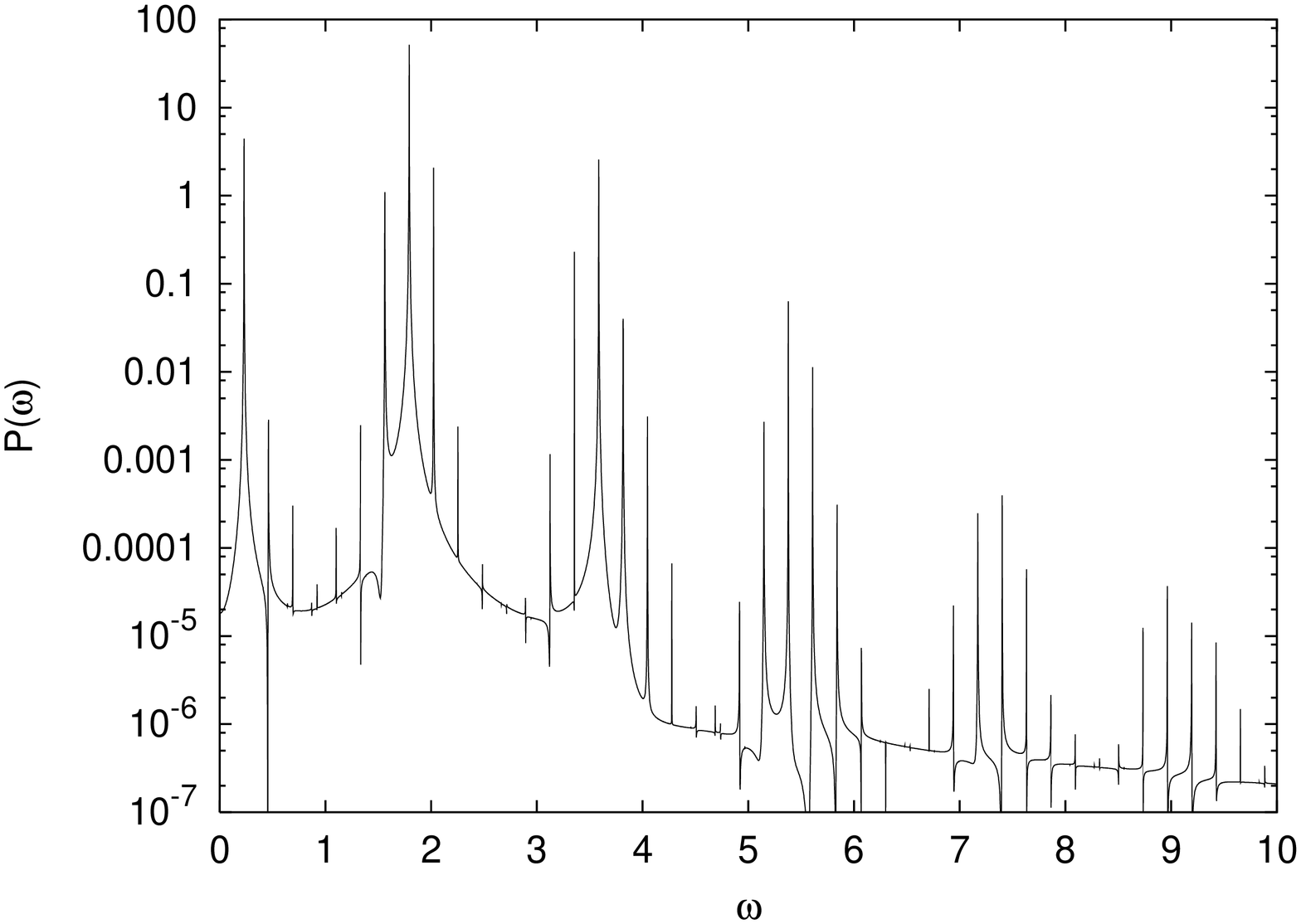}
\caption{Dynamics of a moving ``slow'' soliton for
$Q=0.264$, $C=0.3$ and $g=2.2$
(other parameters same as in Fig.\ \ref{fig:Q0vel}).
Upper figure: intensity  $A_n$ versus time
for 23 consecutive sites (larger $n$ corresponds to later peak time).
Notice that while peaks are equidistant, corresponding to a
well-defined velocity $v\approx 0.285$, the modulation of the peaks is
apparently incommensurate, i.e., no two peaks corresponding to
different sites have identical shape. Lower figure: Fourier spectrum
of the corresponding oscillations of the power $P$ (\ref{norm}),
calculated over 40000 time units. The
main peak at $\omega=2 \pi v \approx 1.793$ corresponds to the translational
motion, while the first peak at $\omega \approx 0.231$ corresponds to
the modulation. All other peaks seen are linear combinations of these.
}
\label{fig:QP}
\end{figure}
%

\begin{figure}[t]
\centering
\includegraphics[height=0.5\textwidth,angle=270]{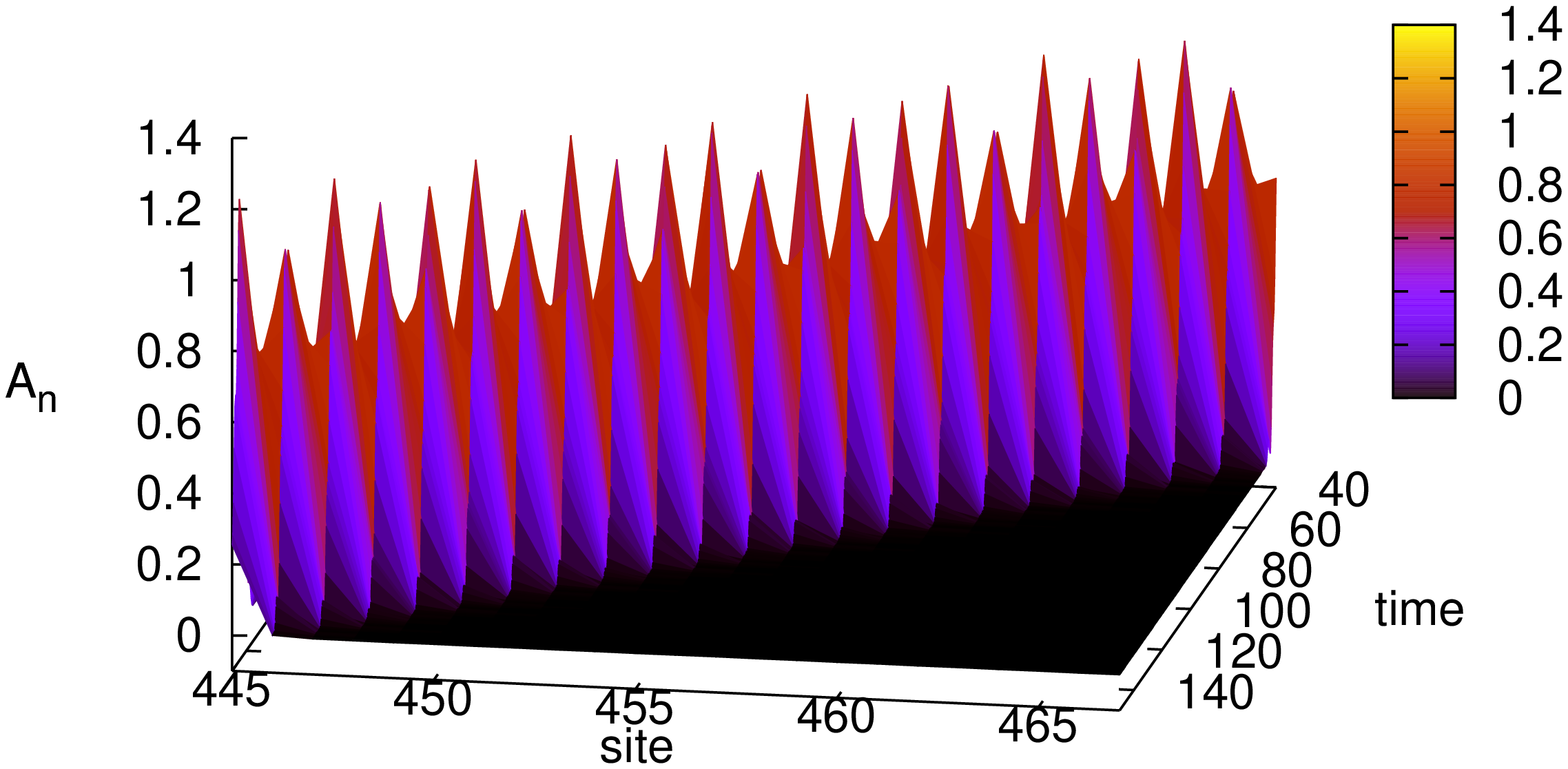}
\includegraphics[height=0.5\textwidth,angle=270]{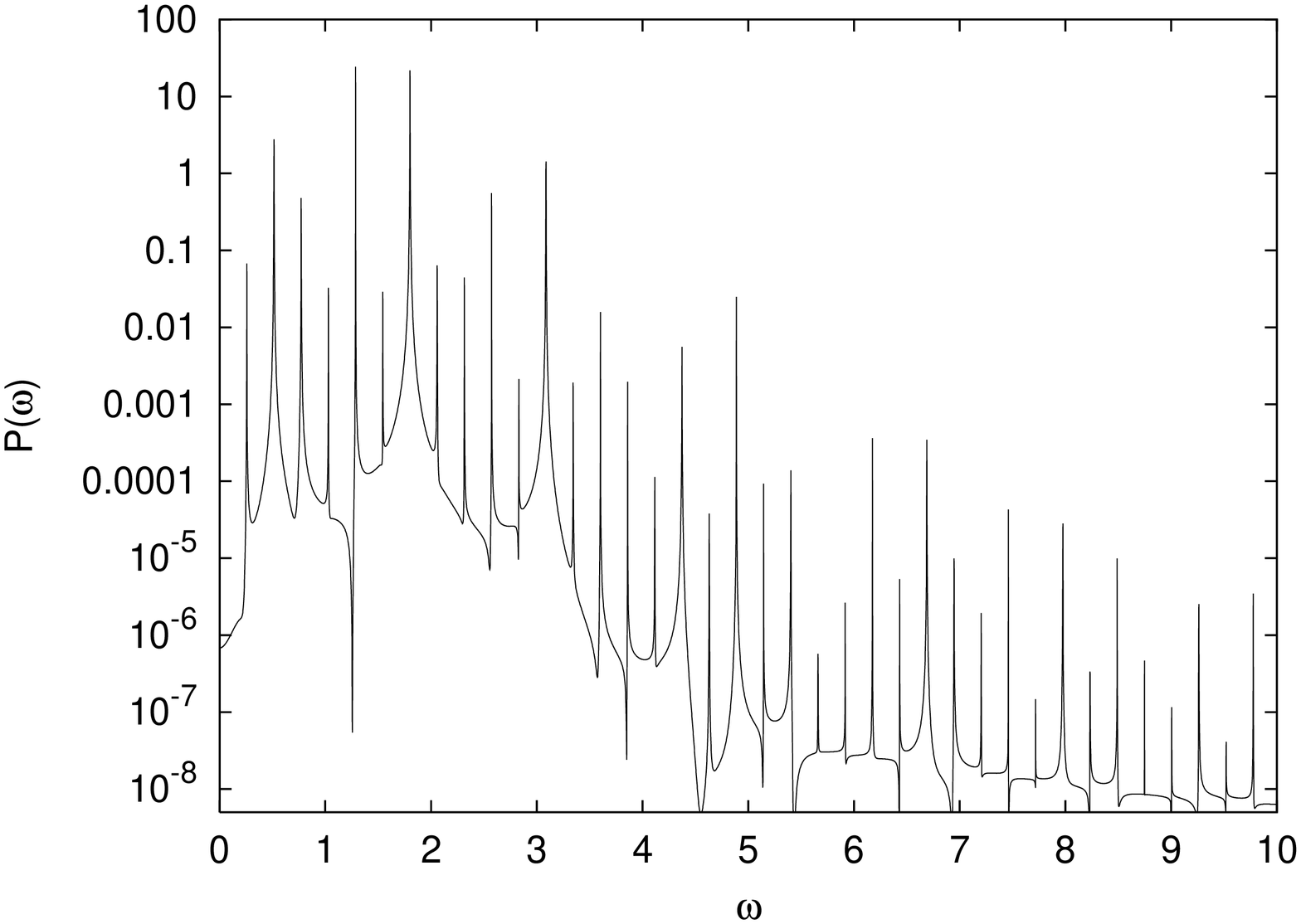}
\caption{(Color online) Dynamics of a soliton moving with velocity
$v \approx 0.205$ in the $(p,q)=(2,5)$
phase-locked regime for
$Q=0.2285$ (other parameters same as in Fig.\ \ref{fig:QP}).
Upper figure: intensity  $A_n$ versus time. Note that the pattern
periodically repeats itself after translating $q=5$ sites.
Lower figure: Fourier spectrum
of the oscillations of the power $P$ (\ref{norm}),
calculated over 40000 time units. The
peak at $\omega_1 = 2 \pi v \approx 1.286$ corresponds to the translational
motion, while the second peak at $\omega_2 \approx 0.514$ corresponds to
the main modulation. The first peak is $\omega_1 - 2 \omega_2 = \omega_2/2$,
and all other peaks seen are multiples of this frequency.
}
\label{fig:2:5}
\end{figure}

Another interesting effect of the weaker pinning potential is the
appearance of parameter regimes where the relation (\ref{pq}) between
internal oscillation frequency and soliton velocity becomes incommensurate.
To illustrate this, we now fix $g=2.2$ and $C=0.3$ and continue the
moving solutions versus $Q$ (from Sec.\ \ref{sec:Q0} we know that no stable
moving solutions exist for these parameter values when $Q=0$). A fast,
stable small-oscillation solution exists for these parameter values only in the
interval $0.264 \lesssim Q \lesssim 0.350$, while the slower solution with
large-amplitude oscillations is found to propagate with a well-defined velocity
down to $Q \approx 0.2095$.  The oscillations and the velocity are
1:1-locked  [$(p,q)=(1,1)$] for $Q\gtrsim 0.265$, where a Hopf-type bifurcation
appears and the oscillations develop an additional modulation frequency,
incommensurate with the translational dynamics as illustrated in
Fig.\ \ref{fig:QP}.

The quasiperiodically moving slow soliton can be found for
$0.224 \lesssim Q\lesssim 0.264$ (note that this corresponds roughly to the
regime where the PN-potential of the corresponding conservative system is
expected to be weakest as discussed above). However, as the modulation
frequency varies (generally increases as $Q$ decreases in this interval),
we also find small intervals of nontrivially phase locked states, where
the two fundamental frequencies lock to rational values (i.e., analogous
to Arnol'd tongues). As an example, we show in Fig.\ \ref{fig:2:5}
the 2:5-locked state [$(p,q)=(2,5)$] which is a stable attractor for
$0.2281 \lesssim Q \lesssim 0.2285$ with other parameters as above.

Decreasing $Q$ further leads to a larger interval of $(p,q)=(1,2)$ locking,
$0.214 \lesssim Q\lesssim 0.227$, which for a further decrease undergoes
a sequence of period-doubling bifurcations leading to a chaotically moving
solution for $Q\lesssim 0.209$, similarly as described above for $Q=0$.
Finally, at $Q \approx 0.15$ it gets trapped into a mode oscillating around
a fixed site, similarly as described in \cite{Floria1,el13}.

\begin{figure}[htbp]
\centering
\includegraphics[height=0.5\textwidth,angle=270]{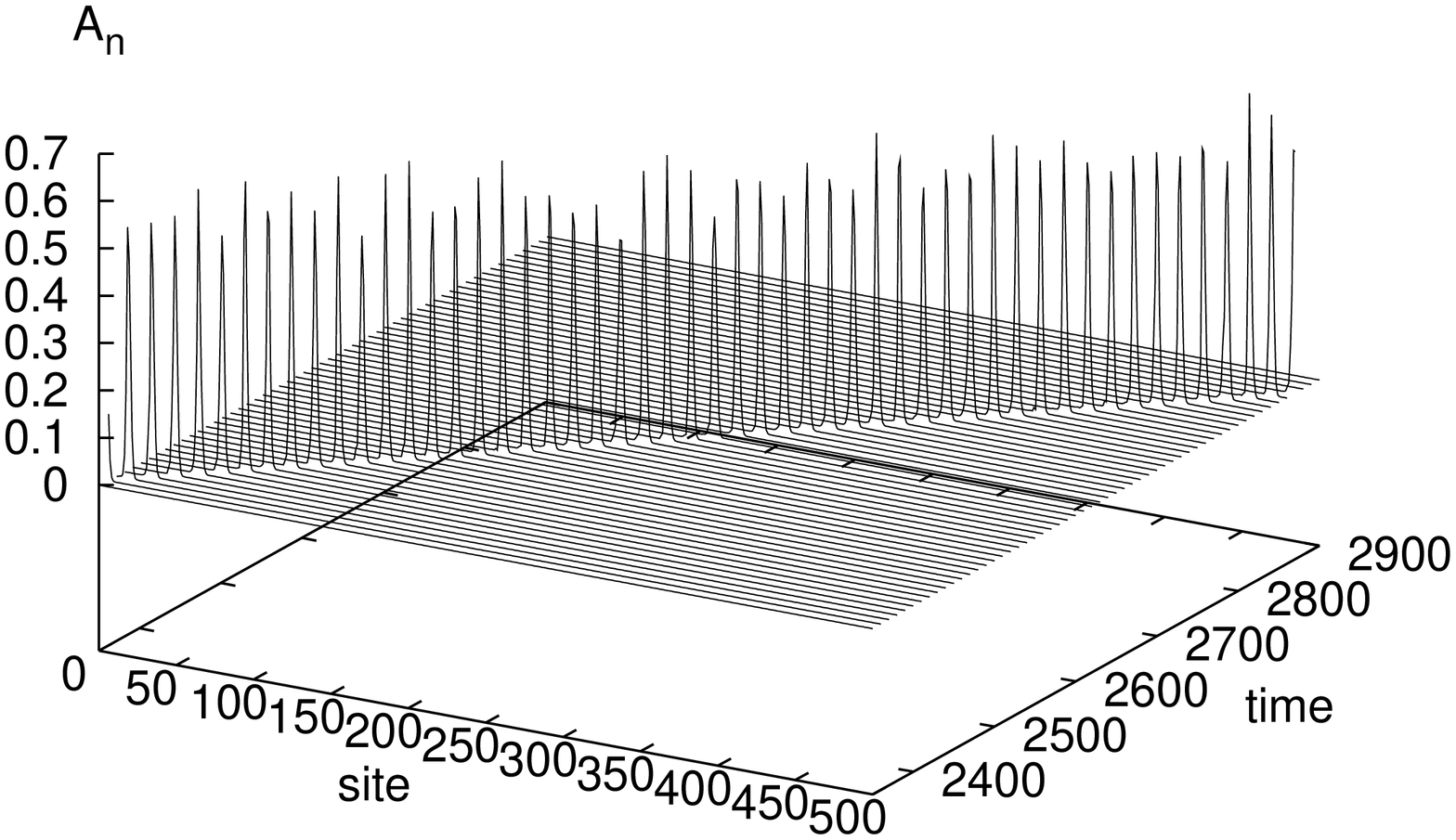}
\includegraphics[height=0.5\textwidth,angle=270]{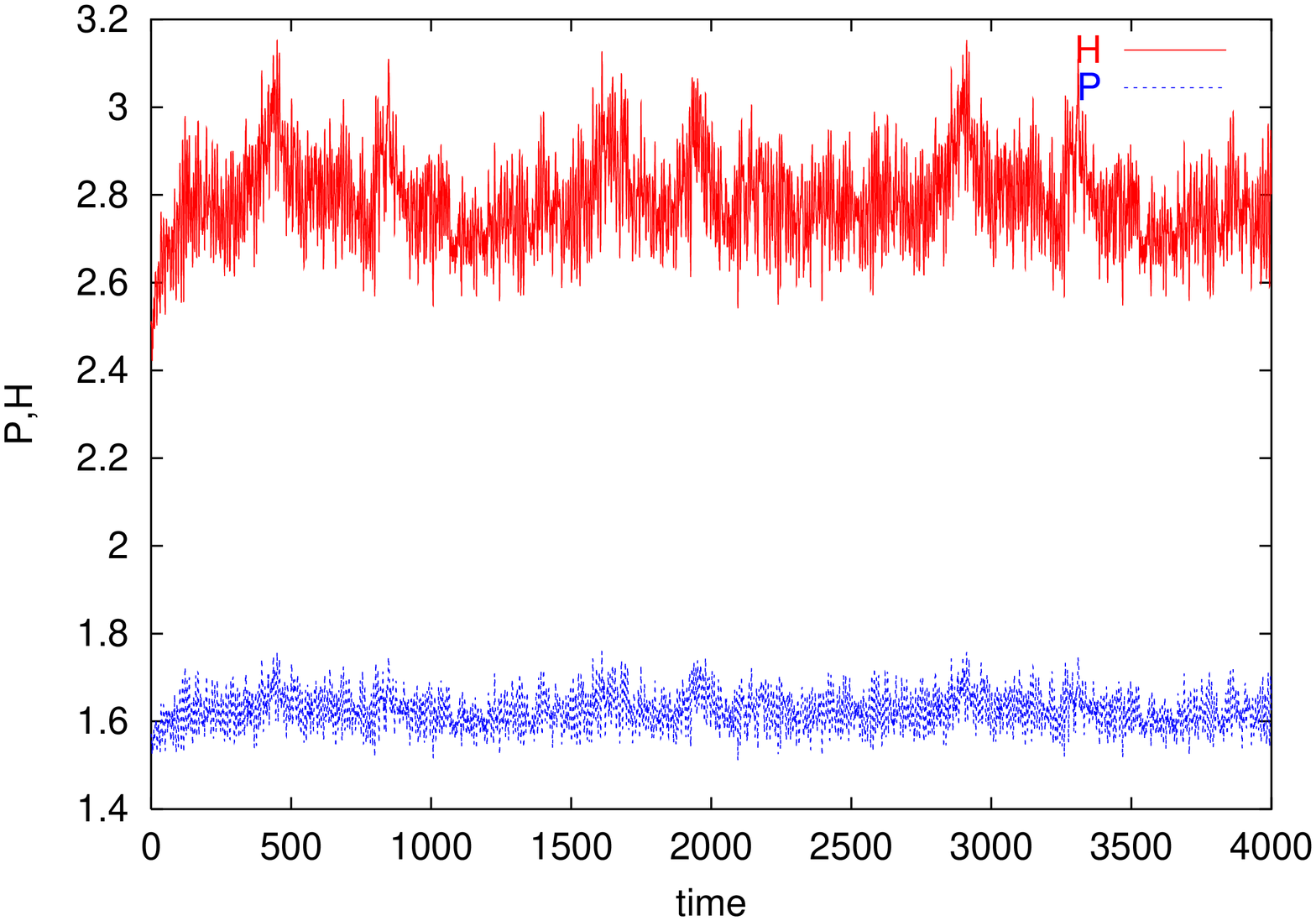}
\caption{(Color online) Dynamics of a soliton moving with velocity
$v \approx 0.977$ in a lattice with uniform disorder of strength $V_0=0.1$,
for $g=2.06$ and other parameters same as in Fig.\ \ref{fig:Q0vel}.
The soliton appears as an attractor using the soliton for $V_0=0$
as initial condition.
Upper figure: intensity  $A_n$ versus time. Only a part of the dynamics
is shown for a lattice with 2405 sites and periodic boundary conditions.
Lower figure: the corresponding oscillations of the quantities
$P$ (\ref{norm}) (lower blue curve) and $H$ (\ref{ham}) (upper red curve).
Note that after an initial transient, the  patterns
periodically repeat themselves after 2462 time units, corresponding to one
round trip in the lattice.
}
\label{fig:dis}
\end{figure}

\subsection{Effects of disorder}
\label{sec:disorder}

Finally, we illustrate that the existence of moving lattice solitons is
not crucially dependent on having a perfect lattice, but also survive in
presence of weak disorder. Generally, disordered photonic structures are of
large current interest, e.g. in the contexts of random lasers and Anderson
localization of light; see, e.g., recent reviews \cite{Wiersma13,ssc13}.
Here, we choose the on-site
potential $V_n$ in
(\ref{master-eq}) to have
a uniform distribution in the interval $[-V_0, V_0]$, and launch initially
a soliton found as an attractor in the regular lattice. If the disorder
is sufficiently weak, the soliton settles into a new attracting moving mode
after a short transient, as illustrated in Fig.\ \ref{fig:dis}. For the
parameter values chosen in  Fig.\ \ref{fig:dis}, which for
$V_0=0$ belong to the branch
of fast solitons for $Q=0$ in Fig.\  \ref{fig:Q0vel}, a moving attracting
state appears for $V_0 \lesssim 0.1$, while the soliton generally gets
trapped for a slightly stronger disorder. The velocity of the soliton in the
disordered lattice in Fig.\ \ref{fig:dis} is
only slightly smaller than for the corresponding
parameter values in the regular lattice ($v\approx 0.977$ compared to
$v\approx 1.092$). From the lower part of Fig.\ \ref{fig:dis}, we see
apparently irregular oscillations of power and Hamiltonian, compensating
for the irregularities in the disordered lattice so that the moving soliton
still can travel with a constant velocity. Note from Fig.\ \ref{fig:dis}
that the oscillation patterns exactly repeat themselves after one round-trip
in the periodic lattice. Thus, even in a disordered lattice the
soliton may at each point adjust its internal parameters according to its
local environment, in order to travel with constant velocity.

As could be expected, the critical disorder below which a traveling mode
can be sustained is generally
considerably smaller for the slow, large-amplitude
oscillation modes. For example, with the parameter values of
Fig.\ \ref{fig:Q0vel}, we found that the moving solution with $g=2.18$ in
Fig.\ \ref{fig:tails} keeps traveling with constant velocity for many round
trips in the lattice when $V_0=0.0006$, but gets trapped for slightly
larger disorder. Also in this case the velocity gets slightly smaller
in presence of disorder ($v\approx 0.059$ compared to
$v\approx 0.062$ for the above parameter values).
\section{Conclusions}
\label{sec:conclusions}
In conclusion, we showed that exact traveling, exponentially localized,
discrete solitons exist as stable dynamical attractors in a model for active
waveguide arrays with Kerr nonlinearity,
supported only by the intrinsic gain and damping, without
explicit external forcing or symmetry-breaking perturbations. The traveling
soliton self-adjusts its internal parameters during the motion
in order to travel with a
constant velocity, leading to oscillations of its total power and Hamiltonian.
In most cases these oscillations are 1:1 locked to the translational motion,
but in certain regimes period-doubling bifurcations to chaotically walking
modes are observed, and likewise Hopf-bifurcations to incommensurate
oscillations and non-trivial phase-locking intervals may appear if also
inter-site Kerr nonlinearities are present. For pure on-site nonlinearity
traveling solitons exist only in a rather narrow parameter regime, which
widens considerably with the inclusion of inter-site nonlinearities, thereby
reducing the Peierls-Nabarro potential obstructing the motion in the
corresponding conservative system. The moving solitons can be divided into
two types, between which bistability appears in certain regimes: fast
modes with small oscillations, and slow modes with large oscillations.
The former may even survive as exact traveling modes if a reasonably strong
disorder is included, while the latter are highly sensitive to trapping.

In our model, the saturable gain and absorption result from the properties
 of the active optical fibers as described in \cite{rosanovbook}. It
is clear that, in this model, all parameter values describing the gain and
damping need to be nonzero
for stable traveling localized solitons to exist; e.g., they would not exist
for purely linear damping, or purely linear gain. Evidently,
it would be highly interesting to understand more precisely what are the
necessary conditions for existence of traveling intrinsic gain-driven
localized modes in more general physical lattice systems, not necessarily
restricted to optics. One particularly interesting example concerns the so
called quodons in layered crystals such as mica, where it has been suggested
that nonlinear lattice excitations could travel for macroscopic distances
by an energy gain resulting from the lattice being in a metastable
configuration \cite{Russell13}. In the optical context, extensions to non-Kerr
nonlinearities and higher dimensions are also relevent, in particularly as 
it is known that saturable \cite{Vicencio} or quadratic \cite{Susanto} 
nonlinearities may considerably 
decrease the Peierls-Nabarro barriers and increase mobility of localized modes 
 also for two-dimensional lattices in the conservative case.

\begin{acknowledgments}
The authors thank A.V.\ Yulin and S.K.\ Turitsyn for useful discussions. 
M.J. also thanks M.\ Marklund for some earlier ideas in this direction. M.J.
thanks the School of Engineering and Applied Science, Aston University, 
for its kind hospitality, and acknowledges
support from the Swedish Research Council. 
J.E.P. appreciates the support of the ERC project ULTRALASER.
\end{acknowledgments}

\end{document}